\documentclass[aps,reprint,10pt,pre,showpacs]{revtex4-1}

\usepackage{amsmath, color}
\usepackage{graphicx}

\draft 

\begin{document}

\title{Crossover Behaviour In Driven Cascades} 

\author{James Burridge}
\email[]{james.burridge@port.ac.uk}

\affiliation{Department of Mathematics, University of Portsmouth, PO1 3HF, United Kingdom.}

\date{\today}

\begin{abstract}
We propose a model which explains how power--law crossover behaviour can arise in a system which is capable of experiencing cascading failure. In our model the susceptibility of the system to cascades is described by a single number, the \emph{propagation power}, which measures the ease with which cascades propagate. Physically, such a number could represent the density of unstable material in a system, its internal connectivity, or the mean susceptibility of its component parts to failure. We assume that the propagation power follows an upward drifting Brownian motion between cascades, and drops discontinuously each time a cascade occurs. Cascades are described by a continuous state branching process with distributional properties determined by the value of the propagation power when they occur. In common with many cascading models, pure power law behaviour is exhibited at a critical level of propagation power, and the mean cascade size diverges. This divergence constrains large systems to the subcritical region. We show that as a result, crossover behaviour appears in the cascade distribution when an average is performed over the distribution of propagation power. We are able to analytically determine the exponents before and after the crossover.

\end{abstract}


\maketitle 

\section{Introduction}
\label{intro}

There are many examples, both man--made and naturally occurring, of phenomena which begin with a small disturbance and end in a catastrophe. Such events include electrical network failures \cite{Dob12}, forest fires \cite{Stau91}, avalanches \cite{Stan96}, nuclear chain reactions \cite{Har64}, snapping ropes \cite{Prad06} and landslides \cite{Pieg06, Herg03,Mal04}. A common feature of the distribution of the sizes of such events, and the mathematical property which links the very small to the very big, is power-law scaling. Such scaling often appears over a number of orders of magnitude with the end of the power law region marked by an exponentially decaying probability density, referred to as an ``exponential cut--off'' \cite{Stau91, Chr92}. The values of the power law exponent, and the cut--off point will depend on the nature of the system, but often not on its fine details -- a phenomenon known as universality \cite{Stau91,Chr92}. The physical origin of the cut--off may be the physical size of the system or its inherent ability propagate the cascade. For infinite systems, there exists a critical level of instability at which the distribution is a pure power-law. As the system approaches this critical point, the cut--off moves increasingly rapidly toward infinity, and the expected cascade size diverges.

In this work it is our aim to investigate the phenomenon of power law crossover in cascade size distributions. Crossover occurs when the distribution of large cascades follows a different power law to that of small cascades.  This type of behaviour was discovered \cite{Prad06} in the distribution of bursts of snapping events in bundles of fibres under tension and close to complete breakdown. It is our aim to show that similar behaviour may be exhibited by a system in which cascades occur repeatedly over an extended period. Each cascade has the effect of increasing the stability of the system so that subsequent cascades propagate less freely. Such a stabilizing effect is seen, for example, in regions prone to forest fires where the extent and frequency of large fires has been reduced by planned burning of limited areas \cite{Boer09}. Another example is where the spread of disease through a population is reduced by the presence of recovered (or vaccinated) individuals, who act as firewalls preventing transmission \cite{Fine11}. We might also expect to see a similar effect in terrain susceptible to landslides. In fact, there is some evidence of crossover behaviour in landslide size distributions \cite{Eeck07}.  However, our model is not tied to one particular physical system, but rather it puts forward a generic explanation of how crossover behaviour might arise. Our approach is to produce the simplest possible model (and explanation) of driven cascades, that is mathematically tractable, and that preserves some key features of real cascading phenomena.

The fundamental quantity of interest to us will be the ability of the system to propagate a cascade, which will be described by a single number, $p$, the propagation power. In physical terms, this quantity might be determined by the connectivity of a network of unstable nodes, the density of unstable material in a system, or the average proximity of the components of a system to failure. Because we view the propagation power as a measure of how easily cascades propagate between parts of system it will not depend on the absolute volume of the system -- it is an intensive property. The dynamical properties of $p$ will be determined by two processes. First, $p$ will tend to increase over time, but with an element of unpredictability, described by Brownian noise \cite{Grim01} so that in the absence of cascades:
\begin{equation}
\label{driving}
dp(t) = \mu dt + \sigma dW(t)
\end{equation}
where $\mu>0$ is the mean rate of increase of $p$, $W(t)$ is a standard Wiener process \cite{Grim01}, and $\sigma>0$ controls the magnitude of the noise. The physical origin of such a process in, for example, a forest fire model might be the drying out of vegetation due to unpredictable weather, or in a landslide model, the natural variability of pore water pressure. We assume that the magnitude of the noise is independent of system size, and therefore it represents an external driving process.

The second influence on $p$ arises from the cascades themselves, which act to stabilize the system. To capture this stabilizing effect, we suppose that if the size of the $k$th cascade since an arbitrary reference time is given by the continuous random variable $J_k$, then in response $p$ changes by $-\epsilon J_k$. The parameter $\epsilon$ measures the sensitivity of $p$ to cascades, and will depend on the size of the system, and the stabilizing effect on those parts of it that are involved in the cascade. For convenience, we will refer to $\epsilon$ as the ``inverse system size''. Note that the distribution of $J_k$ will itself depend on $p$. We define $J(t)$ to be the sum of all cascade sizes that have occurred since $t=0$ and let $N_t$ count the number that have occurred. We then have that:
\begin{equation}
J(t) = \sum_{k=1}^{N_t} J_k.
\end{equation}
Together with the driving dynamics (\ref{driving}) we have a complete stochastic differential equation for $p(t)$:
\begin{equation}
\label{SDE}
dp(t) = \mu dt + \sigma dW(t) + \epsilon\ dJ(t).
\end{equation}
We will assume that cascades begin when a small part of the system spontaneously fails. Since a larger system will have more potentially unstable material then the rate at which cascades begin should scale in proportion to the system size $\epsilon^{-1}$. Assuming that the cascade rate is constant and independent of the cascade history, then the times between cascades will be exponentially distributed and $N_t$ will be a Poisson process with intensity $ \lambda \epsilon^{-1}$, where $\lambda$ is a constant of proportionality, independent of system size. Without loss of generality in the model we can set $\lambda=1$ by re--scaling time and adjusting $\mu$ and $\sigma$.

As we mentioned above, the distribution from which individual cascades sizes, $J_k$, are drawn will depend on the value of $p$ when they occur. We write the probability density function of this distribution $\psi_p(z)$. In section \ref{casc} we introduce a continuous state branching process to describe cascades, which we cap at size $p/\epsilon$ (this cap is proportional to system size, and ensures that the propagation power cannot become negative). We find that:
\begin{equation}
\psi_p(z) \propto z^{-\frac{3}{2}} e^{-\kappa(p) z} \textrm{ for } z< \frac{p}{\epsilon}
\end{equation}
where $\kappa(p) \rightarrow 0$ as $p$ approaches a ``critical value'' $p_c$. At this critical point, $\psi_p(z)$ is a pure power--law and the mean cascade size diverges. For $p<p_c$, $\kappa(p)>0$ so the power--law region is ``cut--off'' at approximately $z \approx \kappa(p)^{-1}$. When $p>p_c$, the mean cascade size increases further still and is infinite in the limit $\epsilon \rightarrow 0$.  The divergence of the mean cascade size as the critical point is approached from below will create a self--stabilizing effect which pushes the system away from $p_c$. The combination of this automatic stabilization and the upward drift of the driving process means that the propagation power will fluctuate about a mean value lying in the subcritical region. We will find that the magnitude of fluctuations is controlled largely by $\sigma$ for large systems.

Of central interest to us is the long term cumulative record of cascade sizes in the system, which we describe with a probability density function $\bar{\psi}(z)$. This will be an average of $\psi_p(z)$ over the values of $p$ at which cascades take place. We define $f(p,t)$ to be the probability density function for the value of the propagation power at time $t$. The ``steady state'' density function for $p$ is then $f(p) := \lim_{t \rightarrow \infty} f(p,t)$. We then have that
\begin{equation}
\label{meanCasc}
\bar{\psi}(z) = \int f(p) \psi_p(z) dp.
\end{equation}
This expression may also be thought of as the probability density function for the size of the next cascade observed after some arbitrary (but large) time. One of our main results is to show analytically that fluctuations in $p$ about its typical subcritical value, causing it to approach temporarily closer to criticality, are what generates crossover behaviour in the ``averaged'' cascade distribution $\bar{\psi}(z)$.

From the above arguments it is clear that typical value of $p$, corresponding to the peak of $f(p)$, will be increased by increasing the driving rate $\mu$. In the limit of infinite system size we will find that the stabilizing effect of diverging mean cascade size near $p_c$ means that it is always the case that $p< p_c$. Therefore, in the limit $\mu \rightarrow \infty$ the system will sit at the critical point. Similar behaviour is exhibited by the Forest Fire model \cite{Dros92}, where the equivalent of $\mu$ is the rate of tree growth. Because there are a wide range of parameter values for which both our model and the Forest Fire model are near critical, both may be seen as examples of ``Self Organised Criticality'' (S.O.C.) \cite{Herg02}. The requirement that $\mu$ be sufficiently large in order for the system to be near criticality means that our model does not exhibit S.O.C. in its purest form, as seen in the Sandpile \cite{Bak88}. However, we can still draw analogy between the steady increase in $p$ and the addition of energy or particles in truly self organising models.

\section{The cascade distribution $\psi_p(z)$}
\label{casc}

Suppose that each cascade begins when a small volume of the system experiences a ``failure'' event (for example it may catch fire, explode, get infected or begin motion). This may induce further parts to fail and so on, forming a sequence of failures whose volumes $X_0, X_1, X_2, \ldots$ are referred to as \emph{generations}. We assume that the generations follow a \emph{continuous state branching process} \cite{Har64,Sen77,Chr92,Stan96}. The relationship between the sizes of the successive generations is encoded by an \emph{offspring} distribution, $G$. This is the probability distribution for the amount of failed material that each unit of the current generation triggers in the next. We assume that $G$ remains the same throughout the cascade.

Suppose that the zeroth generation of the process, $X_0$, is an integer, then the size, $X_1$, of the first generation is the sum of $X_0$ independent copies of a $G$-distributed random variable. In this case the distribution of $X_1$ is simply the convolution of $G$ with itself $X_0$ times, written $G^{\ast X_0}$. If $X_0$ is not an integer, we extend the idea of $n$-fold convolution, following Seneta and Vere--Jones \cite{Sen68}, as follows. Suppose that $Y$ is a random variable drawn from the distribution $G$. The function:
\begin{equation}
\Phi(s) := \mathbf{E}(e^{-s Y})
\end{equation}
is the Laplace transform of $G$. Given this definition, $\Phi(s)^n$ is the Laplace transform of $G^{\ast n}$ where $n \in \{0,1,2,3 \ldots \}$. By relaxing the constraint that $n$ be an integer, we may define the Laplace transform of $X_1$, conditional on $X_0$, to be $\mathbf{E}(e^{-s X_1} \mid X_0) = \Phi(s)^{X_0}$, and then extend this rule to later generations:
\begin{equation}
\label{LapBranch}
\mathbf{E}(e^{-s X_{n+1}} \mid X_n) = \Phi(s)^{X_{n-1}}.
\end{equation}
This recursive equation defines the relationships between the distributions of successive generations, giving a complete characterisation of the process once $G$ is defined.

We will take $G$ to be the Gamma distribution \cite{Dob12,Hus07,Josh06} $\Gamma(k,\theta)$ which has density function
\begin{equation}
\label{GammaDens}
g(x) = \frac{x^{k-1}e^{-\frac{x}{\theta}}}{\Gamma(k) \theta^k},
\end{equation}
defined for $x \geq 0$, where the variables $k>0$ and $\theta>0$ are referred to as the shape and scale parameters. Our choice of $G$ is motivated by the requirement that it have a physically plausible shape (not bimodal), have defined moments, lead to mathematical tractability, and that $X_n \in [0,\infty]$ for all $n \geq 0$. In fact, the asymptotic behaviour of the distribution of the total size of the cascade:
\begin{equation}
Z = \sum_{i=0}^\infty X_i
\end{equation}
will depend only on the mean and variance of the offspring distribution so we expect our crossover predictions would hold for other forms of $G$. A significant advantage of our particular choice is that for all real $n> 0$,  $\Gamma(k,\theta)^{\ast n} \equiv \Gamma(n k,\theta)$, from which we see, via equation (\ref{LapBranch}), that our Gamma branching process is defined by the relationship
\begin{equation}
\label{BP}
X_n \sim \Gamma(k X_{n-1},\theta)
\end{equation}
where $X_0$ is the volume of the first generation of the cascade. In equation (\ref{BP}) the symbol $\sim$ means ``is distributed as''. Using this recursive definition of the cascade, it is possible to compute the distribution of the total cascade size $Z$. However, we first fix the relationship between the offspring distribution and $p$. We will allow this dependence to enter through the mean, $k \theta$, of the offspring distribution, so that the average volume contributed to each successive generation of the cascade per unit volume of the previous generation is proportional to $p$. We choose to let $\theta = p$ and leave $k$ as a free parameter which controls the variance and shape of $G$.

We will write the probability density function for $Z$ as $\psi_p(z)$ and consider first the case where there is no limit on how large cascades can become. We may calculate $\psi_p(z)$ by interpreting $Z$ as the first passage time of a random walk. However, the details of this calculation are somewhat technical, and are not part of our main story, so we confine them to appendix \ref{CSBP}. The resulting probability density function for $Z$ is:
\begin{align}
\label{psi}
\psi_p(z) &= \frac{X_0 p^{-k z} \left(z-X_0\right){}^{k z-1}
   e^{\frac{X_0-z}{p}}}{z \Gamma (k z)} \\
& \sim \left[\sqrt{\frac{k}{2\pi}} X_0 e^{-k X_0 + \frac{X_0}{p }}\right] \frac{ e^{-
\kappa z}}{
   z^{3/2}} \text{ as } z \rightarrow \infty
\label{psiAsymp}
\end{align}
where:
\begin{equation}
\kappa = k \ln (k p) + \frac{1-k p}{p}.
\end{equation}
In the context of equation (\ref{psiAsymp}) the symbol $\sim$ means ``tends asymptotically to''. To the author's knowledge, equation (\ref{psi}) is a new result in the theory of continuous state branching processes \cite{Kyp06}. Considering equation (\ref{psiAsymp}), when the mean of the offspring distribution is one ($kp=1$), then $\kappa=0$, so the distribution is asymptotically a pure $-\tfrac{3}{2}$ power law with infinite moments. The value of propagation power $p_c=\tfrac{1}{k}$ at which this behaviour occurs is referred to as the critical point.  When $p < p_c$ the quantity $\kappa^{-1}$ gives the location of the exponential cut-off.

Provided $k p < 1$, the distribution $\psi_p(z)$ is normalised and its moments are defined. It is useful to have explicit expressions for the first two moments of $\psi_p(z)$ in this case. In appendix \ref{CSBP} we show that, provided the cascade size is not limited by finite system size, then
\begin{align}
\mathbf{E}(Z) &= \frac{X_0}{1-k p} \\
\mathbf{E}(Z^2) &= \frac{X_0^2 (1 -k p )+ X_0 k p
   ^2}{(1-k p)^3}.
\end{align}
When $k p >1$ the distribution (\ref{psi}) is not normalized. This is referred to as the ``supercritical'' regime where infinitely large cascades become possible. The total probability weight is equal to $\mathbf{P}\{Z<\infty\}$. In appendix \ref{CSBP} we show that:
\begin{equation}
\label{norm}
\int_{X_0}^\infty \psi_p(x) dx = e^{\chi(k,p) X_0}
\end{equation}
where:
\begin{equation}
\chi(k, p) = \left(k W_{-1}\left(-\frac{e^{-\frac{1}{k p}}}{k p}\right)+\frac{1}{p}\right) \mathbf{1}_{[\frac{1}{k},\infty]}(p)
\end{equation}
and $W_{-1}$ is one of the two real branches of the Lambert W function \cite{Knu96}, the other being $W_0$.

We will cap cascades at a maximum size
\begin{equation}
Z_{max} = \frac{p}{\epsilon}
\end{equation}
so that they cannot cause the propagation power to take negative values. This cap scales in proportion with the system size. Assuming that cascades will stop abruptly once $Z>Z_{max}$, then the tail of $\psi_p(z)$ will replaced with a delta function $\omega(Z_{max}) \delta(z-Z_{max})$ where $\omega(Z_{max})$ is the total probability weight in the tail, plus the probability of an infinite cascade:
\begin{equation}
\omega(Z_{max}) = \int_{Z_{max}}^\infty \psi_p(z) dz + (1-e^{\chi(k,p) X_0 }).
\end{equation}
The capped cascade distribution is therefore:
\begin{equation}
\psi_p(z ;Z_{max}) := \mathbf{1}_{[X_0,Z_{max}]}(z) \psi_p(z) + \omega(Z_{max}) \delta(z-Z_{max}),
\end{equation}
and the $n$th cascade moment in a finite system is:
\begin{equation}
\label{mom}
\mathbf{E}(Z^n) :=  \int_{X_0}^{Z_{max}} z^n \psi_p(z) dz + \omega(Z_{max})Z_{max}^n,
\end{equation}
where the second term arises from integration over the delta function in the capped cascade density. For simplicity, for the remainder of the paper we will begin all cascades with a failure of volume $X_0=1$.

\section{The distribution of propagation power}

\begin{figure}
\includegraphics[width=7.5 cm]{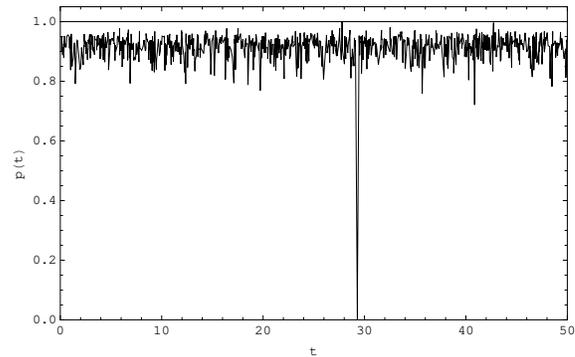}
\caption{\label{Sim} Simulation of $p(t)$ using parameter values are $\epsilon=10^{-5}$, $\mu=10$ and $\sigma=0.5$ and $k=1$ giving a critical $p$ value of $p_c=1.0$. Values of $p(t)$ were recorded when the jump which preceded them exceeded $1\%$ of the system size.   }
\end{figure}

For arbitrary $\epsilon$, the steady state distribution of the propagation power equation (\ref{SDE}) cannot be found analytically, but may be determined by simulation. However, for small $\epsilon$, jumps which are not small with respect to the system size become increasingly rare, and the dynamics of $p$ may be approximated by diffusion with drift, giving rise to a pure diffusion equation which is analytically tractable. In the limit $\epsilon \rightarrow 0$, this approximation becomes exact. To illustrate the rarity of large jumps, Figure \ref{Sim} shows a simulation of $p(t)$ when $\epsilon=10^{-5}$, $\mu=10$, $\sigma=0.5$ and $k=1.0$. In this case out of $0.5 \times 10^7$ cascades only six exceeded $10\%$ of the system size, and the system reached criticality only once out of all recorded times.

To derive our diffusion approximation we view $p$ as evolving in discrete steps, with its value being recorded immediately before each cascade. Each change in $p$ is therefore comprised of a cascade, followed by a period of diffusion until the next cascade occurs. This defines a discrete time stochastic process $p_1, p_2, p_3 \ldots$ where $p_n$ is the propagation power immediately before the the $n$th cascade. We let $f_n(p)$ be the probability density function for $p_n$. We then define $W_y(r)$ to be the probability density function for the size, $r$, of the next single step of the process given that the current value of propagation power is $y$. Note that $W_y(r)$ is independent of $n$. The master equation governing the evolution of $f_n(p)$ is then:
\begin{equation}
\label{master}
f_{n+1}(p) = \int  W_{p-r}(r)f_n(p-r) dr.
\end{equation}
Our aim is to find the steady state distribution: $\lim_{n \rightarrow \infty} f_n(p)$.

Before we continue our analysis of the master equation we will make clear the link between the distributions of discrete time process, and the underlying continuous time process (\ref{SDE}). In section \ref{intro}, we defined $f(p)$ to be the steady state probability density function for the continuous process, $p(t)$, which may intuitively be thought of as the density function for the value of $p$ observed at an arbitrary time $T$, large enough so that the influence of initial conditions is insignificant. Due to the properties of exponential waiting times \cite{Law06}, the time interval, $\Delta T$, between $T$ and the last cascade before $T$ will be exponentially distributed. Letting $p^*$ be the value of $p$ immediately after this last cascade, we see that $p(T) = p^\ast + \mu \Delta T + \sigma W(\Delta T)$. Now let $T'$ be another large time, but restricted to the set of cascade times so that we are making an observation of the discrete time process. Again using the properties of exponential waiting times, the time $\Delta T'$ since the previous cascade will have the same distribution as $\Delta T$ (intuitively this arises because $T$ is likely to lie in a larger than average waiting interval). Because the value of $p$ just after the previous cascade will be drawn from the same distribution as $p^\ast$, we have that $p(T) =^d p(T')$ where $=^d$ denotes equality in distribution. As a result, observations of the discrete time and the continuous time process have the same distribution at large times:
\begin{equation}
\lim_{n \rightarrow \infty} f_n(p) =  f(p).
\end{equation}
If the times between cascades were not exponentially distributed then the probability distribution of the state of the system immediately preceding a cascade would not, in general, be the same as the distribution at a randomly selected time. In that case equation (\ref{meanCasc}) would be incorrect because $\psi_p(z)$ must be averaged over the distribution of $p$ immediately preceding a cascade.

To approximate the steady state solution to (\ref{master}) we derive the corresponding Kramers--Moyal equation \cite{Kamp} by expanding the integrand of (\ref{master}) in powers of $r$:
\begin{multline*}
W_{p-r}f_n(p-r) \approx W_p(r)f_n(p) - r \partial_p\{W_p(r)f_n(p)\} \\
+ \frac{r^2}{2} \partial_{pp}\{W_p(r)f_n(p)\} + \ldots
\end{multline*}
Substituting this approximation back into the master equation we find that:
\begin{multline*}
f_{n+1}(p) = f_n(p) -\partial_p \left(f_n(p) \int r W_p(r) dr \right) \\
+\frac{1}{2} \partial_{pp} \left(f_n(p) \int r^2 W_p(r) dr \right) + \ldots
\end{multline*}
where we have made use of the normalisation of the step size distribution to simplify the first term. We now define the first two moments of step size to be:
\begin{align}
A(p) &:= \int r W_p(r) dr \\
B(p) &:= \int r^2 W_p(r) dr.
\end{align}
As $\epsilon \rightarrow 0$, both the time between cascades and their effect on $p$ decline so the step distribution becomes increasingly sharply peaked about $r=0$. We therefore ignore moments of higher order than two and obtain a discrete time analogue of the Fokker--Planck equation for pure diffusion:
\begin{equation}
\label{FP}
f_{n+1}(p) - f_n(p) = -\partial_p \{A(p) f_n(p)\} + \frac{1}{2} \partial_{pp}\{B(p) f_n(p)\}.
\end{equation}
We are interested in the steady state behaviour of the process, where $f_{n+1}(p)=f_n(p)\equiv f(p)$, and in particular we require the function $f(p)$, which from equation (\ref{FP}), satisfies:
\begin{equation}
\label{J}
\partial_p \left[A(p)f(p)\right] = \frac{1}{2} \partial_{pp}\left[B(p)f(p)\right]
\end{equation}
The first integral of this equation, the probability current $A(p)f(p) - \tfrac{1}{2} \partial_{p}(B(p)f(p))$, will be zero.

In the limit of large system size, for $p < p_c$, the probability of cascades which engulf the entire system is zero, and the system can never access the region $p \geq p_c$ because the mean cascade size diverges in the neighborhood of $p=p_c$. We therefore expect our diffusion approximation to become exact. As this limit is approached our expressions for $A(p)$ and $B(p)$ take the asymptotic form:
\begin{align}
A(p) &\sim  \epsilon \left[\mu - \mathbf{E}(Z)\right] \\
B(p) &\sim  \epsilon \sigma^2 + \epsilon^2 \left[\mathbf{E}(Z^2) - 2\mu \mathbf{E}(Z) +2 \mu^2\right]
\label{BAsymp}
\end{align}
In equation (\ref{BAsymp}) we have retained the $\epsilon^2$ term because the moments of the cascade distribution become very large as $p \rightarrow p_c$. For any finite system we expect our approximation to break down near this critical point, and to breakdown globally if there is sufficient probability weight in the supercritical region where infinite cascades have nonzero probability. We will explore this breakdown using simulations.  The first moment of the cascade distribution has asymptotic behaviour:
\begin{equation}
\mathbf{E}(Z) \sim \frac{1}{1-k p} \text{ as } \epsilon \rightarrow 0
\end{equation}
which diverges near $p=p_c=\tfrac{1}{k}$, creating an infinite negative drift. The probability weight at the critical point therefore declines to zero as $\epsilon \rightarrow 0$ and the divergence in $B(p)$ will never be realised. We may therefore drop the $\epsilon^2$ term for infinite systems. Making use of $A(p)$ and the simplified  $B(p)$ we see from equation (\ref{J}) that the limiting form of $f(p)$ satisfies:
\begin{equation}
\label{fEqn}
\left(\mu-\frac{1}{1-kp}\right)f(p) = \frac{\sigma^2}{2} f'(p).
\end{equation}
Although the cascade distribution $\psi_p(z)$ is not defined for $p<0$, in the interests of tractability, we will take equation (\ref{fEqn}) as valid over the interval $[-\infty,0]$, yielding the following expression for $f(p)$:
\begin{equation}
\label{f_pdf}
f(p) = \frac{k \mu \left(\frac{2\mu}{k \sigma^2}\right)^{\frac{2}{k \sigma^2}}}{\Gamma\left[\frac{2}{k \sigma^2}\right]} (1-k p)^{\frac{2}{k \sigma^2}} \exp\left[-\frac{2 \mu}{k \sigma^2}(1-kp)\right]
\end{equation}
For all parameter values of interest to us, the probability weight in the invalid region $p \in [-\infty,0]$ is less than $10^{-12}$. Note that this solution is independent of $\epsilon$, but we expect it to become an increasingly good approximation to the true solution as $\epsilon \rightarrow 0$ for values of $p < p_c$.

\section{The steady state cascade size distribution $\bar{\psi}(z)$}

\begin{figure}
\includegraphics[width=8 cm]{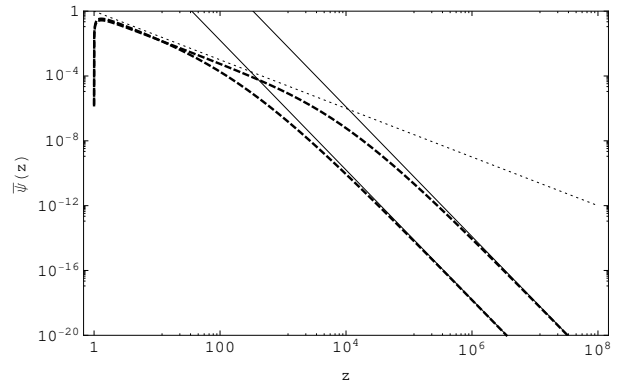}
\caption{\label{mu} Log--log plot of the mean cascade distributions (dashed lines) $\bar{\psi}(z)$ in the limit $\epsilon \rightarrow 0$ when $k=2$, $\sigma=0.5$ and $\mu$ takes the values $10$ and $60$. The larger $\mu$ value produces a crossover point at a higher value of $z$. Also shown are the asymptotic power--law predictions (\ref{asymp}) (solid lines) and the function $z^{-\frac{3}{2}}$ as a dotted line.  }
\end{figure}

\begin{figure}
\includegraphics[width=8 cm]{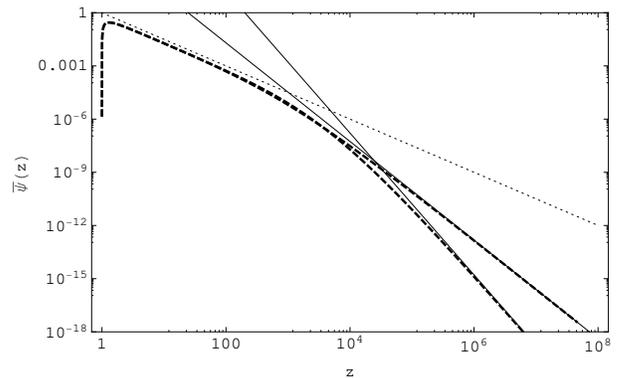}
\caption{\label{sig} Log--log plot of the mean cascade distributions (dashed lines) $\bar{\psi}(z)$ in the limit $\epsilon \rightarrow 0$ when $k=2$, $\mu=40$ and $\sigma$ takes the values $0.5$ and $0.8$. The larger $\sigma$ value produces a shallower tail in the cascade distribution. Also shown are the asymptotic power--law predictions (\ref{asymp}) (solid lines) and the function $z^{-\frac{3}{2}}$ as a dotted line.   }
\end{figure}

By making use of the large system size approximation to the propagation power probability density, $f(p)$, together with the asymptotic form of the cascade distribution (\ref{psiAsymp}), is it possible to investigate the $(z \rightarrow \infty)$ asymptotic behaviour of the mean cascade distribution in the limit $\epsilon \rightarrow 0$. Throughout this section all asymptotic formulae hold as $z \rightarrow \infty$. To simplify our expressions we define a new parameter:
\begin{equation}
\alpha := \frac{2}{k \sigma^2}.
\end{equation}
The mean cascade distribution may then be written
\begin{align}
\bar{\psi}(z)& =  \int f(p) \psi_p(z) dp \\
 & \sim \frac{ k^{\frac{3}{2}} e^{-k} \mathcal{C}}{\sqrt{2}} z^{-\frac{3}{2}} \int_{-\infty}^{p_c} e^{\frac{1}{p}-\kappa z} (1-kp)^\alpha e^{- \alpha \mu (1-kp)} dp
\label{psiBar}
\end{align}
where
\begin{equation}
\mathcal{C} = \frac{ \mu (\alpha \mu)^\alpha }{\sqrt{\pi} \Gamma(\alpha)}.
\end{equation}
We have extended the lower limit of integration to $-\infty$ for tractability, since the integrand will be negligible when $p<0$. The integral (\ref{psiBar}) is not tractable. However, as $z \rightarrow \infty$, the weight of the integrand becomes concentrated in a shrinking neighborhood of the critical point. We may therefore approximate the integral asymptotically by replacing the first exponent in the integrand with its Taylor expansion to quadratic order about $p_c=\tfrac{1}{k}$:
\begin{equation}
\frac{1}{p} - \kappa  z \approx k\left(1 + (1-k p) - \frac{1}{2}(z-2)(1-k p)^2\right).
\end{equation}
Making the change of variables $s=1-kp$, our approximation becomes:
\begin{equation}
\bar{\psi}(z) \sim \mathcal{C} \sqrt{\frac{k}{2}}  z^{-\frac{3}{2}} \int_0^\infty  s^\alpha e^{-\left(\alpha \mu-k\right)s -\frac{k}{2}(z-2) s^2} ds.
\end{equation}
We now make a second change of variables:
\begin{equation}
(z-2)s^2 = t^2,
\end{equation}
which gives:
\begin{multline}
\label{para}
\bar{\psi}(z) \sim  \mathcal{C} \sqrt{\frac{k}{2}} z^{-\frac{3}{2}} \left(\frac{2}{k(z-2)}\right)^{\frac{1+\alpha}{2}} \\ \times \int_0^\infty  t^\alpha e^{-t^2} \exp\left[-\left(\frac{2}{k(z-2)}\right)^{\frac{1}{2}} (\alpha \mu-k)t\right] dt.
\end{multline}
We may extract the asymptotic behaviour of this integral by noting that, for finite $t$, as $z \rightarrow \infty$ the $z$ dependent exponential term tends to one. We now note that provided $\alpha>1$:
\begin{equation}
\int_0^\infty  t^\alpha e^{-t^2} dt = \frac{1}{2} \Gamma \left[ \frac{1+\alpha}{2} \right].
\end{equation}
The asymptotic behaviour of the mean cascade distribution is therefore a pure power law:
\begin{equation}
\label{asymp}
\bar{\psi}(z) \sim  \mathcal{A} z^{-2-\frac{1}{k \sigma^2}}
\end{equation}
where
\begin{equation}
\mathcal{A} = \frac{1}{2} \left(\frac{2}{k}\right)^{\frac{\alpha}{2}}   \Gamma \left[ \frac{1+\alpha}{2} \right]\mathcal{C}.
\end{equation}
So, in the limit of large system size, the tail of the mean cascade distribution, rather than being exponential, follows a power law with an exponent which is an increasing function of the variance of the destabilisation process. For smaller $z$, when most of the probability weight in $f(p)$ corresponds to cut-offs at larger $z$ values, we see  $z^{-\frac{3}{2}}$ behaviour preserved. Together with our asymptotic predictions this gives rise to a power--law crossover. The theoretical distribution $\bar{\psi}(z)$, in the limit $\epsilon \rightarrow 0$, is illustrated in Figures \ref{mu} and \ref{sig}  together with the asymptotic result (\ref{asymp}).

\section{Simulation Results}

Using simulations we will now test the validity of our large system crossover predictions, explore the influence of  $\epsilon$ and the effect of the model parameters on the averaged cascade distribution, $\bar{\psi}(z)$.

\subsection{Techniques of simulation}

We make use of two simulation methods; a ``naive'' technique where every cascade is simulated, and an ``accelerated'' technique which uses a diffusion approximation when the probability of  jumps of significant size in comparison to $\epsilon^{-1}$ is sufficiently small.

\subsubsection{Naive simulation}

\begin{figure}
\includegraphics[width=8 cm]{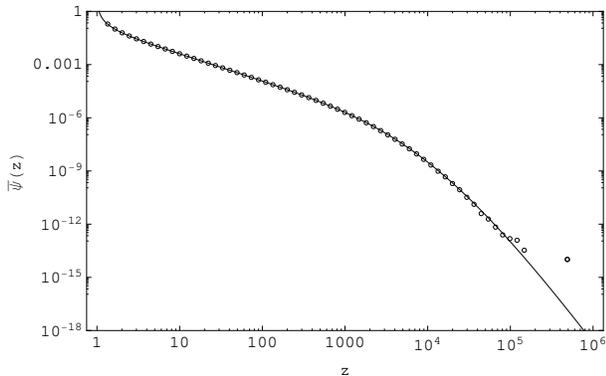}
\caption{\label{naive} The circles show the simulated cascade distribution $\bar{\psi}(z)$ when $\epsilon=10^{-6}$, $\mu=10$, $\sigma=1.5$ and $k=0.1$. The results were obtained by simulating the propagation power process over $10^9$ cascades. The black line shows the theoretical cascade distribution in the limit $\epsilon \rightarrow 0$.    }
\end{figure}

The simplest method to determine the averaged cascade distribution is to simulate the stochastic process described by equation (\ref{master}). The results (shown in Figure \ref{naive}) were obtained by simulating $10^9$ cascades, and recording their sizes in bins of increasing width. It is clear from the figure that to fully examine the tail behaviour of the distribution would require us to simulate significantly more cascades. This would be prohibitively time consuming so we adopt an accelerated scheme for the remainder of our results.

\subsubsection{Accelerated simulation}

For the parameters used in Figure \ref{naive} we see that our theoretical cascade distribution in the limit $\epsilon \rightarrow 0$ was accurate for cascades up to at least $\tfrac{1}{100}$th of the system size. Due to the exponential cut-off in $\psi_p(z)$, large cascades are only likely when $p$ is in the vicinity of the critical point. The behaviour of the distribution of $p$ in this region determines the tail behaviour of the average cascade distribution. Given that the diffusion approximation works well for smaller $p$ values, we may speed up our simulation by explicitly simulating cascades only once we are close to $p_c$. For lower $p$ values we approximate the process as pure diffusion, which is much faster to simulate. We determine the switch point $p_s$ by requiring that for $p<p_s$, the probability of observing at least one cascade that is larger than $\tfrac{1}{100}$th the system size over the simulation must be effectively zero (taken to be $10^{-12}$). For $p<p_s$ we advance $p(t)$ using a fixed time step $\delta t$. For $p>p_s$ we simulate every cascade, where the times between them are exponentially distributed with mean $\epsilon$. Typically, $\delta t$ may be taken to be orders of magnitude larger than $\epsilon$, but must be small enough so that the process doesn't significantly ``overshoot'' $p_s$ into the region where the quality of the diffusion approximation is uncertain. We refer to the sum total of long ($\delta t$) and short (mean length $\epsilon$) time steps to be the number of steps in the simulation. Provided $\delta t >> \epsilon$, then, because the system spends most of its time in the region $p<p_s$, we can effectively simulate $p(t)$ over a much larger time interval than with the naive method. Rather then generating a catalogue of cascades, this technique produces an approximate density function for $p$. We average the cascade distribution over this approximate density function in order to find $\bar{\psi}(p)$.

\begin{figure}
\includegraphics[width=8 cm]{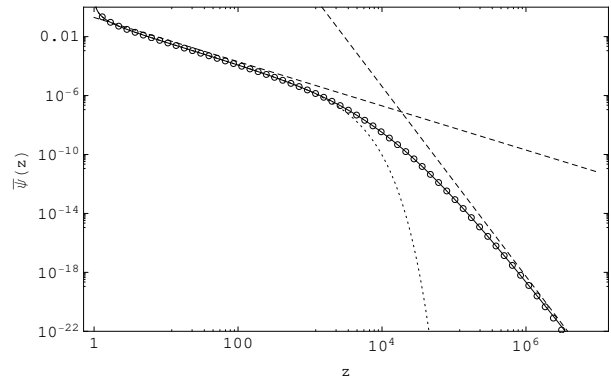}
\caption{\label{accel} The circles show the simulated cascade distribution $\bar{\psi}(z)$ when $\epsilon=10^{-9}$, $\mu=10$, $\sigma=1.5$ and $k=0.1$. The results were obtained by simulating the propagation power process using the accelerated technique over $10^9$ time steps. The black line shows the theoretical cascade distribution in the limit $\epsilon \rightarrow 0$. The dashed lines are pure power laws. The shallow gradient line has exponent $-\tfrac{3}{2}$, whereas the steep gradient line is our asymptotic prediction (\ref{asymp}). The dotted line shows $\psi_p(z)$ when $p$ is equal to its mean value. Note the presence of the exponential cut--off.}
\end{figure}

In Figure \ref{accel} we have used this accelerated method to investigate the cascade distribution for a larger system than in Figure \ref{naive} but with otherwise identical parameter values. Also plotted is our asymptotic prediction (\ref{asymp}), and the cascade density $\psi_p(z)$ when $p$ is equal to its mean value. This serves to highlight the difference between crossover behaviour and exponential cut--off. We note that the analytic predictions are well matched in this case; the diffusion approximation remained valid some distance beyond the switch point $p_s$.

\subsection{Validity of large system crossover predictions}

\begin{figure}
\includegraphics[width=8 cm]{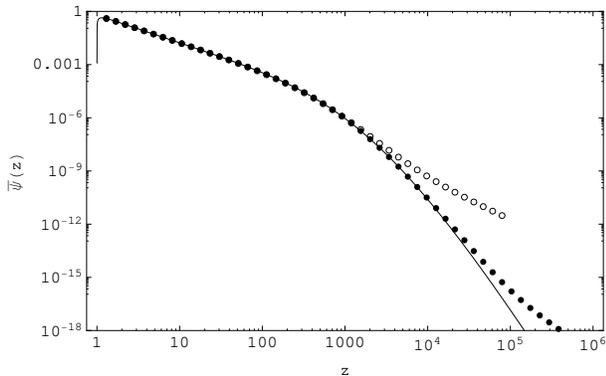}
\caption{\label{breakdown} The open circles show the simulated cascade distribution $\bar{\psi}(z)$ when $\epsilon=10^{-5}$, $\mu=15$, $\sigma=0.35$ and $k=1.5$. The filled circles show the simulated cascade distribution when $\epsilon=10^{-7}$ and all other parameters are the same. The results were obtained by simulating the propagation power process using the accelerated technique over $10^9$ time steps. The solid line shows the theoretical cascade distribution in the limit $\epsilon \rightarrow 0$. }
\end{figure}

We now investigate the breakdown of our crossover predications by simulating smaller systems. In Figure \ref{breakdown} we have simulated two different sized systems that are both smaller than in Figure \ref{accel}. We note that when $k>1$ the distribution of the first generation of the cascade possesses a maximum located away from zero and the cascade distribution inherits this characteristic. For the smaller system, the crossover fails to fully develop, and the exponent begins to increase again. This occurs because the distribution of propagation power does not decay to zero at the critical point, so the averaged cascade distribution includes significant contributions from values of $p$ for which $\psi_p(z)$ is approximately a pure power law for $z<\epsilon^{-5}$. The breakdown of the diffusion approximation occurs in part because the frequency of cascades is insufficient to realise their divergent mean size on short time scales. In the larger system where $\epsilon = 10^{-7}$ we see that the crossover develops more fully. Because cascades occur with greater frequency, fluctuations in the short term average of the cascade size are reduced.

\subsection{The influence of $\mu$ and $\sigma$}

In Figures \ref{mu} and \ref{sig}, we illustrate the role of the parameters $\mu$ and $\sigma$ in determining the behaviour of the cascade distribution $\bar{\psi}(z)$ in the limit of large system size.  Figure \ref {mu} shows that for fixed $\sigma$, the driving rate $\mu$, which determines the location of the maximum of $f(p)$, fixes the location of the crossover. At larger driving rates, the crossover point shifts to larger cascade sizes. This is because, for larger $\mu$, the peak of $f(p)$ is nearer to the critical point. Therefore the typical exponential cut--off scale is larger and the $-\tfrac{3}{2}$ scaling region is extended.

In Figure \ref{sig} we see that a noisier driving process reduces the magnitude of the tail exponent, but the size of the $-\tfrac{3}{2}$ region remains unaffected. Increased variability in the driving process means that although the typical distance from the critical point is unchanged, the system spends more time in close proximity to it and therefore large cut--offs are more heavily weighted.

In both Figures the asymptotic power--law predictions (\ref{asymp}) for the cascade distribution give an indication of the size of the region over which the crossover occurs.

\section{Conclusions}

We have presented a driven cascade model which, in the limit of large system size, exhibits power--law crossover behaviour in its cascade size distribution. For smaller systems, the crossover partially develops, but the distribution moves back toward the initial $-\tfrac{3}{2}$ power law at larger cascade sizes, because the system is able to reach and exceed the critical value of propagation power.

The mechanism which generates the crossover is a competition between the driving process, which increases the instability of the system, and the cascade process, which reduces it. As the propagation power nears the critical point, $p_c$, the mean cascade size diverges, so that in the limit of large system size, $p_c$ is not accessible for finite $\mu$. The upper tail of the distribution of propagation power, $f(p)$, therefore decays to zero at $p_c$. The asymptotic behaviour of the averaged cascade distribution, $\bar{\psi}(z)$, is determined by averaging the asymptotic behaviour of the cascade distribution for fixed propagation power, $\psi_p(z)$, over the upper tail of $f(p)$. The result is a power--law with exponent lower than $-\tfrac{3}{2}$, producing a crossover.

We suggest that the presence of a crossover in a cascade size distribution may indicate that the system is able to self stabilize through frequent but non--catastrophic cascades, and that the mechanism which drives the instability of the system to cascading failure is inherently noisy. The stabilizing cascades act to prevent the system from reaching a fully critical state where very large cascades can occur. Because our system will be near critical for a wide range of parameter values (large enough $\mu$), it may be considered to exhibit  ``Self Organised Criticality''. However, the crossover is more obvious if the typical value of propagation power is not too close to criticality, and it has been our focus to explore crossover. It should also be noted that an infinite system will only become fully critical in the limit $\mu \rightarrow \infty$, whereas a finite system may become critical or supercritical through fluctuations.  The location of the crossover indicates how close to criticality the system will typically be found, because it is determined by the peak of $f(p)$. It remains to adapt the ideas contained in our simple model to investigate real physical systems such as forest fires and landslides.


\appendix

\section{Cascade distribution for Gamma branching process}

\label{CSBP}

The purpose of this appendix is to derive the probability density function for the total cascade size in the continuous state branching process defined by relationship (\ref{BP}), and also to determine the probability of an infinite size cascade.

\subsubsection{Mapping to a first passage problem}

We begin by showing that the our problem may be interpreted as a first passage time problem. Consider the branching process $X_0, X_1, X_2, \ldots$ defined by the relationship $X_n \sim \Gamma(k X_{n-1},\theta)$ with $X_0$ given.  In order to calculate the distribution of $Z = \sum_{k=0}^\infty X_k$, we show that $Z$ may be viewed as the first passage time of a stochastic process through the origin. We first give the definition of the ``Gamma process'' \cite{Kyp06}, which takes place in continuous time on $[0,\infty]$. If $S_t$ is a Gamma process with parameters $k$ and $\theta$ then:
\begin{enumerate}
\item
$S_0 = 0$.
\item
It has independent increments, in the sense that for any $0 \leq t_0 < t_1 < \ldots < t_n$ the random variables $S_{t_1}-S_{t_{0}}, S_{t_2}-S_{t_1}, \ldots S_{t_n}-S_{t_{n-1}}$ are independent and:
\item
$S_{s+t} - S_s \sim \Gamma(k t, \theta)$.
\end{enumerate}
We now define a new stochastic process $Q_t = X_0 + S_t -t$, where $X_0$ is the size of the first generation of our branching process. By defining $Z_n := \sum_{k=0}^n X_k$ to be the cumulative cascade size up to the $n$th step, we may show that the processes $Q_t$ and $X_n$ have the following relationship:
\begin{equation}
\label{map}
X_{n+1} =^d Q_{Z_n}
\end{equation}
provided that the cascade has not ended for some $k<n$. Here $=^d$ denotes equality in distribution. We may deduce that this relationship holds inductively. We note first that from the defining properties of the Gamma process:
\begin{equation}
S_{Z_n}-S_{Z_{n-1}} \sim \Gamma(k X_n,\theta) \sim X_{n+1}.
\end{equation}
Assuming that relationship (\ref{map}) holds for all $n<k$ then
\begin{align}
Q_{Z_k} &= Q_{Z_{k-1}} + (S_{Z_k}-S_{Z_{k-1}})-(Z_k-Z_{k-1}) \\
    &=^d X_k +  X_{k+1} - X_k \\
    &= X_{k+1}.
\end{align}
Since $Q_{Z_0} = X_0 + S_{X_0}-X_0 =^d X_1$ then (\ref{map}) holds when $n=1$ and therefore for all $n$ by induction. The cascade ends at the first generation for which $X_n=0$, at which point $Q_{Z_{n-1}}=0$, so the total size of the cascade is equal to the first time that the process $Q_t$ meets the origin. In order to find the cascade size distribution we need to solve this first passage time problem.

\subsubsection{Gamma process as the limit of a discrete process}

Our first passage time problem is most easily solved by viewing the stochastic process $Q_t$ as the limit of a discrete state random walk. Here we show how the appropriate random walk is constructed.

We begin by noting that the negative binomial distribution, which has probability mass function:
\begin{equation}
b(n,r,q) = \frac{\Gamma(n+r)}{n! \Gamma(r)} (1-q)^r q^n
\end{equation}
provides an arbitrarily close discrete approximation to the gamma distribution for appropriate choice of the parameters $r$ and $q$. The approximation is set up in the following way. We divide $[0,\infty]$ into a discrete lattice of constant spacing $\delta$, and let $X_{\delta}$ be a discrete random variable which approximates $X \sim \Gamma(k,\theta)$. Let $X_{\delta}$ have the probability mass function:
\begin{equation}
\mathbf{P}(X_\delta=n \delta) = b(n,r,q).
\end{equation}
The mean and variance of $X_\delta$ are then $\mathbf{E}(X_\delta) = \tfrac{\delta q r}{1-q}$ and $\text{Var}(X_\delta)= \tfrac{\delta^2 q r}{(1-q)^2}$. Letting $r^\ast$ and $q^\ast$ be the values of $r$ and $q$ for which the mean and variance of $X$ and $X_\delta$ are equal, we find that:
\begin{align}
r^\ast &= \frac{k\theta}{\theta-\delta}  \\
q^\ast &= 1-\frac{\delta}{\theta}.
\end{align}
With these choices of $r$ and $q$, in the limit $\delta \rightarrow 0$ the discrete distribution converges to $\Gamma(k,\theta)$ in the following sense:
\begin{equation}
\lim_{\delta \rightarrow 0} \frac{1}{\delta} b \left[\lfloor x /\delta \rfloor,r^\ast,q^\ast \right] = \frac{x^{k-1}e^{-\frac{x}{\theta}}}{\Gamma(k) \theta^k},
\end{equation}
where the notation $\lfloor a \rfloor$ represents the integer part of the real number $a$. As well as providing an arbitrarily good approximation to the Gamma distribution, the negative binomial distribution also has the property that if $Y \sim NB(r,q)$ is a negative binomial variable, then the sum of $m$ independent copies of $Y$ has distribution $NB(mr,p)$. This allows us to set up a discrete approximation to the Gamma process as follows. We divide each unit of time into $\delta^{-1}$ subunits and let the random variable $A$ have distribution:
\begin{equation}
A \sim NB(\delta r^\ast, q^\ast).
\end{equation}
We will refer to this as the ``atomic'' random variable with ``atomic distribution''. We may approximate the Gamma process as a sum of a sequence, $A_1, A_2, \ldots $ of independent copies of $A$:
\begin{equation}
S_t \approx \delta \sum_{k=1}^{\lfloor t /\delta \rfloor} A_k.
\end{equation}
In other words, we are viewing $S_t$ as a scaled random walk who's step sizes are given by the atomic distribution. We may approximate the process $Q_t$ in a similar way to $S_t$:
\begin{equation}
\label{QWalk}
Q_t \approx \delta \left(\lfloor  X_0/\delta \rfloor + \sum_{k=1}^{\lfloor t /\delta \rfloor} A_k - \lfloor  t/\delta \rfloor \right).
\end{equation}

\subsubsection{First passage time of the discrete process}

Having shown how to construct the discrete state random walk, we now solve the first passage problem using generating functions \cite{Har64} and the Lagrange inversion formula \cite{Wilf05}.

Let $m = \lfloor X_0 /\delta \rfloor$ and $Z_\delta(m)$ be the first passage time of the walk (\ref{QWalk}), starting from position $m$. Considering the first step, which will have size $A-1$, we have that:
\begin{equation}
\label{rec}
Z_\delta(m) = 1 + Z_\delta(m+A-1).
\end{equation}
Since the time for the walk to get from position $m$ to the origin is equal to the time it takes to get to position $1$ plus the time to get from position $1$ to the origin, then  $Z_\delta(m) =^d Z_\delta(m-1) + Z_\delta(1) $. The quantity $Z_\delta(m)$ is therefore the sum of $m$ independent copies of $Z_\delta(1)$. We have from (\ref{rec}) that:
\begin{equation}
\label{ZRec}
Z_\delta(1) = 1 + Z_\delta(A).
\end{equation}
If $H(s)$ and $F(s)$ are the probability generating functions for $Z_\delta(1)$ and $A$, then from equation (\ref{ZRec}) we have
\begin{align}
H(s) &= \mathbf{E}(s^{1 + Z_\delta(A)}) \\
&= s \mathbf{E}[ \mathbf{E}(s^{Z_\delta(A)}\mid A) ] \\
&= s \mathbf{E}[ (H(s))^A ] \\
&= s F(H(s)).
\end{align}
From the negative binomial mass function we have that:
\begin{equation}
F(s) = \sum_{n=0}^\infty s^n b(n,\delta r^*,q^*) = \left(\frac{1-q^*}{1-q^* s}\right)^{\delta r^*}.
\end{equation}
We are interested in the probability generating function of $Z_\delta(m)$, which is just $H^m(s)$. The coefficient of $s^n$ in this function may be determined using the Lagrange inversion formula \cite{Wilf05}:
\begin{align}
[s^n] H^m(s) &= \frac{1}{n}[H^{n-1}]\left\{ \left(\frac{d}{dH} H^{m}\right) F^n(H)\right\} \\
&= \frac{m}{n} [H^{n-m}] F^n(H) \\
&= \frac{m}{n} \frac{\Gamma(n(1+\delta r^*)-m)}{\Gamma(n\delta r^*) \Gamma(n-m+1)}(1-q^*)^{\delta r^*n} (q^*)^{n-m} \\
&= \mathbf{P}\{Z_\delta(m) = n\}
\end{align}
where the notation $[x^n]f(x)$ stands for the coefficient of $x^n$ in the Taylor series of $f(x)$. We now have the probability mass function for the cascade size in the discrete branching process which approximates the continuum process that we are interested in.

\subsubsection{Continuum limit of the discrete process}

Now that we have the solution to the discrete problem we solve the continuous problem by taking the continuum limit. We find expressions for the moments of the cascade size using a similar method. Using a martingale method we determine the probability that the cascade is of finite size in the supercritical case.

We obtain the continuum cascade density function, which we will call $\psi(z)$, by setting $n=z/\delta$ and $m=X_0/\delta$ and then taking the limit $\delta \rightarrow 0$ of $\mathbf{P}\{Z_\delta(m) = n\}$:
\begin{align}
\psi(z) &= \lim_{\delta \rightarrow 0} \frac {1}{\delta} \mathbf{P}\{Z_\delta(m) = n\} \\
&= \frac{X_0 \theta^{-k z} \left(z-X_0\right){}^{k z-1}
   e^{\frac{X_0-z}{\theta}}}{z \Gamma (k z)}.
\label{exact}
\end{align}
The asymptotic properties of $\psi(z)$ may be determined by making use of Stirling's approximation: $\Gamma(z+1) \sim \sqrt{2 \pi z} \left(\tfrac{z}{e}\right)^{z}$. The result is:
\begin{equation}
\label{cDist}
\psi(z) \sim \left[\sqrt{\frac{k}{2\pi}} X_0 e^{-k X_0 + \frac{X_0}{\theta }}\right] \frac{ e^{-
\kappa z}}{
   z^{3/2}} \text{ as } z \rightarrow \infty
\end{equation}
where
\begin{equation}
\kappa = k \ln (k \theta) + \frac{1-k \theta}{\theta}.
\end{equation}

Provided $k \theta < 1$, the distribution $\psi(z)$ is normalised and its moments are defined. It is useful to have explicit expressions for the first two moments of $\psi(z)$ in this case. We may compute the moments of the (discrete) distribution of $Z_\delta(1)$ by differentiating the generating function relationship: $H(s)=sF(H(s))$, and then solving for $H'(s)$ and $H''(s)$. Using the expression for $F(s)$, together with the fact that when $k \theta <1$, $H(1)=F(1)=1$, we find that:
\begin{align}
\mathbf{E}(Z_\delta(1)) &=  \frac{\delta}{1-k \theta} \\
\text{Var}(Z_\delta(1)) &= \frac{\delta k \theta^2}{(1-k \theta)^3}.
\end{align}
 The total cascade size, $Z$, has the same distribution as the sum of $X_0 \delta^{-1}$ copies of $Z_\delta(1)$, so $\mathbf{E}(Z) = X_0 \delta^{-1} \mathbf{E}(Z_\delta(1))$ and $\text{Var}(Z^2) =X_0 \delta^{-1} \textrm{Var}(Z_\delta(1))$, yielding the first two moments of the cascade distribution in exact form:
\begin{align}
\mathbf{E}(Z) &= \frac{X_0}{1-kp} \\
\mathbf{E}(Z^2) &= \frac{X_0^2 (1 -k \theta )+ X_0 k \theta
   ^2}{(1-k \theta)^3}.
\end{align}
Numerical integration of the exact distribution (\ref{exact}) reveals that it is not normalized when $k \theta >1$. This is the ``supercritical'' regime. In general the total probability weight is equal to $\mathbf{P}\{Z<\infty\}$, which is less than one in the supercritical case because there is a non--zero probability of seeing an infinite cascade. We may deduce this probability by considering the stochastic process:
\begin{equation}
M_t = e^{-q Q_t}.
\end{equation}
Taking the expectation value of this process, conditional on its value at $t=0$ we find that:
\begin{equation}
\mathbf{E}(M_t) = e^{-q X_0 + t(q- k \ln(1+q \theta))}.
\end{equation}
If we let $q$ be the solution to the equation $q- k \ln(1+q \theta)=0$ then this expectation will be independent of time. The value of $q$ which solves this equation is:
\begin{equation}
q^\ast = -\frac{1}{\theta} - k W_{-1}\left(-\frac{1}{k \theta} e^{-\frac{1}{k \theta}}\right) \geq 0.
\end{equation}
Letting $Z$ be the first time at which the process meets the origin, then we have that \begin{equation}
\mathbf{E}(M_Z) =  \mathbf{P}(Z < \infty) = e^{-q^\ast X_0}.
\end{equation}
This gives the result presented in equation (\ref{mom}).

%

\begin{acknowledgments}
The author would like to acknowledge Murad Banaji, Samia Burridge, Alexey Kuznetsov, Andreas Kyprianou and Malcolm Whitworth for useful discussions, as well as the anonymous referees for careful reading and constructive criticism.
\end{acknowledgments}


\begin{thebibliography}{99}

\bibitem{Dob12}{
J. Kim, K. R. Wierzbicki, I. Dobson and R. C. Hardiman, IEEE Systems Journal \textbf{26} (3), 548 (2012)}

\bibitem{Stau91}{
D. Stauffer and A. Aharony, \emph{Introduction to Percolation Theory} (CRC Press, 1991).}

\bibitem{Stan96}{
K. B. Lauritsen, S. Zapperi, and H. E. Stanley,  Phys. Rev. E \textbf{54} 2483 (1996).}

\bibitem{Har64}{
T. E. Harris, \emph{The Theory of Branching Processes} (RAND Corporation 1964).}

\bibitem{Prad06}{
S. Pradhan, A.  Hansen and P. C. Hemmer,  Phys. Rev. E \textbf{74} 016122 (2006).}

\bibitem{Pieg06}{
E. Piegari, V. Cataudella, R. Di Maio, L. Milano and M. Nicodemi, Phys. Rev. E \textbf{73}, 026123  (2006).}

\bibitem{Herg03}{
S. Hergarten,  Natural Hazards and Earth System Sciences \textbf{3} 505 (2003)}

\bibitem{Mal04}{
B. Malamud, D. L.  Turcotte, F. Guzzetti and  P. Reichenbach, Earth Surface Processes and Landforms \textbf{29} 687 (2004).}

\bibitem{Chr92}{
K. Christensen, Ph.D. thesis, University of Aarhus, 1992.}


\bibitem{Sen77}
{C C Heyde and E Seneta, \emph{I J Bienaymé : Statistical theory anticipated} (Springer-Verlag, New York-Heidelberg, 1977)}


\bibitem{Boer09}
{M. M. Boer,  R. J. Sadler, R. S. Wittkuhn, L. McCaw, and P. F. Grierson, Forest Ecology and Management
\textbf{259}(1), 132 (2009).}

\bibitem{Eeck07}
{M. Van Den Eeckhaut, J. Poesen, G. Govers, G. Verstraeten and A. Demoulin, Earth. Planet. Sci. Lett. \textbf{256} 588 (2007)}




\bibitem{Grim01}
{G. Grimmett and D. Stirzaker, \emph{Probability and Random Processes} (Oxford University Press, 2001)}


\bibitem{Fine11}
{P. Fine, K. Eames and D. L. Haymann, Clinical Infectious Diseases \textbf{52} (7), 911 (2011)}

\bibitem{Dros92}{
B. Drossel and F. Schwabl, Phys. Rev. Lett. \textbf{69}, 1629 (1992).}

\bibitem{Herg02}{
S. Hergarten, \emph{Self--Organized Criticality in Earth Systems} (Springer, 2001).}

\bibitem{Bak88}{
P. Bak, C. Tang and K. Wiesenfeld, Phys. Rev. A. \textbf{38} 364 (1988).}

\bibitem{Sen68}{
E. Seneta and D. Vere--Jones, Z. Wahrscheinlichkeitstheorie und verw. Gebiete \textbf{10}, 212 (1968).}

\bibitem{Hus07}
{G. J. Husak, J. Michaelsen and C. Funk, International Journal of Climatology \textbf{27} (7) 935 (2007)}

\bibitem{Josh06}
{M. S. Joshi and A. M. Stacey, Risk Magazine, July, 78 (2006)}

\bibitem{Knu96}
{R. M. Corless, G. H. Gonnet, D. E. G. Hare, D. J. Jeffrey and D. E. Knuth, Advances in Computational Mathematics 329 (1996)}


\bibitem{Law06}
{G. F. Lawler,  \emph{Introduction to Stochastic Processes} (Chapman \& Hall, London New York, 2006)}

\bibitem{Kamp}{
N. G. Van Kampen,  \emph{Stochastic Processes in Physics and Chemistry} (Elsevier, 2007).}


\bibitem{Abr65}
{M. Abramowitz and I. A. Stegun \emph{Handbook of Mathematical Functions} (Dover, New York, 1965)}

\bibitem{Kyp06}
{A. E. Kyprianou, \emph{Introductory Lectures on Fluctuations of Levy Processes with Applications} (Springer-Verlag, Berlin Heidelberg New York, 2006)}

\bibitem{Wilf05}
{
H. Wilf, \emph{Generatingfunctionology} (A K Peters/CRC Press,  2005)}

\end{thebibliography}

\end{document}